\documentclass[final,2p,times]{elsarticle}
\usepackage{amsmath}
\usepackage[usenames]{color}
\definecolor{blue}{rgb}{0,0.08,0.45}

\usepackage{amssymb}

\newcommand{\ud}[1]{#1^\dagger}

\journal{Superlattices and Microstructures}

\begin{document}

\begin{frontmatter}



\title{Effective cavity pumping\\from weakly coupled quantum dots}


\author{E.~del Valle}
\author{F.~P. Laussy}

\address{School of Physics and Astronomy, University of Southampton, Southampton, UK}

\begin{abstract}
  We derive the effective cavity pumping and decay rates for the
  master equation of a quantum dot-microcavity system in presence of
  $N$ weakly coupled dots. We show that the in-flow of photons is not
  linked to the out-flow by thermal equilibrium relationships.
\end{abstract}


\end{frontmatter}

Strong coupling between a single quantum dot (QD) an a microcavity
mode was achieved for the first time in
2004~\cite{reithmaier04a,yoshie04a} and has since witnessed many
technical progresses, from hitting the quantum
limit~\cite{hennessy07a} and quantum nonlinearities~\cite{kasprzak10a}
to lasing~\cite{nomura10a} and scalable
implementation~\cite{dousse09a}. Prospects for this field are great
given the ever increasing quality of structures and reaching better
strong coupling (see \cite{laussy10b} for a review). On the
theoretical side of this fundamental physics, we have shown that
including the incoherent continuous excitation, typical of these
experiments, is essential to reproduce quantitatively the observed
spectral anticrossing~\cite{laussy08a}. Non-vanishing pumping also
affects the structure of dressed
states~\cite{delvalle09a,arxiv_delvalle10a} as compared to its
spontaneous emission counterpart. The direct and continuous excitonic
pumping is provided by the income of electron-hole pairs optically
generated or electrically injected in the wetting
layer~\cite{averkiev09a}. However, also some cavity pumping is
necessary in order to successfully fit the spectral
lineshapes~\cite{laussy08a,laucht09b,munch09a}. Different mechanisms
may produce some income of cavity photons, such as the well known
thermal excitation~\cite{cirac91a} or the cascade de-excitation of
multiexciton states~\cite{winger09a}. It has been recently argued that
only the thermal type of excitation is physically admissible for the
cavity~\cite{ridolfo10a,yao10a}.  This was refuted in
Ref.~\cite{comment09a} based on physical arguments. Here, we show that
a simple model can make explicit a case where cavity pumping is not,
indeed, of a thermal character. The model describes the
situation---appealing on physical grounds---where the dot that
strongly couples to the cavity mode, is surrounded by several
``spectator'' dots that are in weak-coupling. These also get excited
by the excitonic pumping, and emit preferentially, due to Purcell
enhancement, in the cavity mode.

We thus consider an assembly of QDs with Fermion lowering
operators~$\sigma_i$ ($i$ labelling the $i$th dot), each coupled with
strength~$g_i$ to the single Boson mode~$a$ of a microcavity. The QDs
are detuned by a small quantity~$\Delta_i$ from the cavity mode, with
the rotating wave Hamiltonian:
$H_i=\Delta_i\ud{\sigma_i}\sigma_i+g_i(\ud{a}\sigma_i+a\ud{\sigma_i})$. Each
dot is further endowed with dissipation (at rate $\gamma_i$) and
excitonic pumping (at rate $P_i$) in the Lindblad form, $\mathcal{L}_i
\rho=\frac{\gamma_i}2(2\sigma_i\rho\ud{\sigma_i}-\ud{\sigma_i}\sigma_i\rho-\rho\ud{\sigma_i}\sigma_i)+\frac{P_i}2(2\ud{\sigma_i}\rho\sigma_i-\sigma_i\ud{\sigma_i}i\rho-i\rho\sigma_i\ud{\sigma_i})$. The
cavity dissipation (at rate $\gamma_a$) and cavity pumping (at rate
$P_a$) also are in the Lindblad form, $\mathcal{L}_a
\rho=\frac{\gamma_a}2(2a\rho\ud{a}-\ud{a}a\rho-\rho\ud{a}a)+\frac{P_a}2(2\ud{a}\rho
a-a\ud{a}\rho-\rho a\ud{a})$. The total density matrix
operator~$\rho$, follows the master equation:
\begin{equation}
  \label{eq:MonApr5173226BST2010}
  \partial_t\rho=\sum_i\Big( i[\rho,H_i ]+\mathcal{L}_i\rho\Big)+\mathcal{L}_a\rho\,.
\end{equation}
If the dots are uncoupled to the cavity, $g_i=0$, we can easily obtain
from the rate equations
$\partial_t\langle\ud{a}a\rangle=-\gamma_a\langle\ud{a}a\rangle+P_a(1+\langle\ud{a}a\rangle)$
and
$ \partial_t\langle\ud{\sigma_i}\sigma_i\rangle=-\gamma_i\langle\ud{\sigma_i}\sigma_i\rangle+P_i(1-\langle\ud{\sigma_i}\sigma_i\rangle)$,
the steady state populations for cavity mode and dots:
\begin{equation}
  \label{eq:FriOct2111525BST2009}
  n_a\equiv \langle\ud{a}a\rangle=P_a/\Gamma_a\,,\quad n_i\equiv\langle\ud{\sigma_i}\sigma_i\rangle=P_i/\Gamma_i\,.
\end{equation}
They are given in terms of the bosonic and fermionic effective
decoherence rates:
\begin{equation}
  \label{eq:FriOct2141538BST2009}
  \Gamma_a\equiv \gamma_a-P_a\,,\quad \Gamma_i\equiv \gamma_i+P_i\,,
\end{equation}
which are also the uncoupled spectral linewidths of cavity and dots,
as obtained through the quantum regression
formula~\cite{delvalle_book10a}. The number of photons can be
arbitrarily large while each QD takes (average) values between 0
and~1.  If the coupling is weak or the detuning to the cavity mode is
large, the populations and linewidths are still given by
expressions~(\ref{eq:FriOct2111525BST2009}) and
(\ref{eq:FriOct2141538BST2009}), but for some effective decay and pump
parameters, $\gamma^\mathrm{eff}$, $P^\mathrm{eff}$, as shown below.

All $\Gamma$s and~$P$s are as yet essentially undefined
phenomenological parameters. They could be linked in some way, e.g.,
if the dissipation arises from a thermal bath, the following well
known relationship would link them (in both cavity and QD cases):
\begin{equation}
  \label{eq:FriOct2104449BST2009}
  \gamma=\kappa(\bar n_T+1)\,,\quad P=\kappa\bar n_T\,.
\end{equation}
$\kappa$ is the zero temperature decay rate of the mode and $\bar n_T$
is the mean number of excitations in the reservoir at temperature $T$.
This leads to thermal equilibrium average populations $n_a=\bar n_T$
and $n_i=\bar n_T/(2 \bar n_T +1)$ and naturally prohibits features of
out-of-equilibrium, such as inversion of population for the dot (here
$n_i$ remains below $1/2$) or line narrowing for the cavity
($\Gamma_a=\kappa_a$ remains a constant, independent of $T$), however
high the occupancy $\bar n_T$ of the thermal mode. A thermal bath is a
medium of \emph{loss}, as $\gamma>P$ (as seen clearly in
Eq.~(\ref{eq:FriOct2104449BST2009})). In out-of-equilibrium
conditions, especially under an externally applied pumping, one can
expect deviations from the thermal scenario. We will see in what
follows a simple model of a \emph{gain} medium for the cavity, in the
sense that the linewidth decreases while the effective pumping rate
increases, with the external excitation.

In the bosonic case, if~$P_a$ and~$\gamma_a$ are allowed to vary
independently, there is an obvious singularity in
Eq.~(\ref{eq:FriOct2141538BST2009}) at~$P_a=\gamma_a$, past which
point values are negative. This is because, although the master
equation is still valid at all finite time, it has no steady state.
The physical reason why, is clear enough: more particles are injected
at all time than are lost by decay. Therefore populations increase
without bound (they diverge in the infinite times).  There is no
deeper physics here than the fact that no all dynamical systems have a
steady state, some because they are oscillatory, others because they
increase without bounds.  The general consideration of pump and decay
bosonic rates, finds its most important domain of applicability with
atom lasers and polariton
lasers~\cite{holland96a,imamoglu96b,porras03a,rubo03a,
  laussy04c,schwendimann06a,schwendimann08a,doan08a}, that is, systems
where a condensate (or coherent state) is formed by scattering of
bosons into the final state from another state rather than by
emission. In both cases, scattering or emission, the process is
stimulated. In this case the income and outcome of particles is a
complicated function of the distribution of excitons (or polaritons)
in the higher $k$-states (see, e.g., Ref~\cite{imamoglu96b}) and even
the case $\gamma_a(t)<P_a(t)$ can then be realized in the
transients~\cite{laussy04b}.

On the other hand, in the fermionic (QD) case, there are no
divergences, and parameters $P_i$ and~$\gamma_i$ are, in general,
considered as independent in the literature. In fact, within the
theory of the single-atom laser, a more general relationship between
the emitter pump and decay rates, is broadly used~\cite{briegel93a}:
$\gamma_i=\Gamma_i(1-s_i)$, $\quad P_i=\Gamma_i s_i$, where $s_i$ is
limited to the interval $[0,1]$. This form now describes
when~$s_i>1/2$, \emph{gain} of the QD from the reservoir which leads
to its population inversion $n_i>1/2$. This can be theoretically
mapped to a thermal bath with negative
temperature~\cite{gardiner_book00a}.

We consider that among all the dots in the sample, only one, say
$i=0$, couples strongly ($g_0\gg \gamma_a$) and resonantly
($\Delta_0=0$) to the cavity mode. The dynamics of strong coupling
between this dot and the cavity, the so-called \emph{Jaynes-Cummings}
model, cannot be described perturbatively and must in general be
solved to all orders and numerically~\cite{delvalle09a}. We further
investigate the rest of the dots in the sample, $i=1,\hdots N$, that
are weakly coupled or/and far from resonance to the cavity mode. Their
effect on the cavity population and spectral properties is
perturbative (Purcell effect) and can be considered, as a first
approximation, independent of the strong coupling physics with dot
$0$. We, therefore, solve Eq.~(\ref{eq:MonApr5173226BST2010}) with
$i>0$, to first order and then trace out the weakly coupled dots
degrees of freedom. This will result in some effective parameters
$\gamma_a^\mathrm{eff}$ and $P_a^\mathrm{eff}$ that are to appear
eventually in the final reduced master equation, that was the starting
point in our earlier work:
\begin{equation}
  \label{eq:WedApr7184035BST2010}
  \partial_t\rho=i[\rho,H_0]+\mathcal{L}_0\rho+\frac{\gamma_a^\mathrm{eff}}2(2a\rho\ud{a}-\ud{a}a\rho-\rho\ud{a}a)+\frac{P_a^\mathrm{eff}}2(2\ud{a}\rho
a-a\ud{a}\rho-\rho a\ud{a})\,.
\end{equation}
The line broadening of the cavity mode neglecting the strongly-coupled
dot (when $g_0=0$) is given by
$\Gamma_a^\mathrm{eff}=\gamma_a^\mathrm{eff}-P_a^\mathrm{eff}$.

We start by solving the dynamics of the cavity with only one of the
weakly coupled dots: $i=1$. Only one-photon correlations need be
considered, which is equivalent to solving the dynamics truncating in
the first rung~\cite{laussy09a,delvalle09a}. The solutions are
analytical, $n_a=P_a^\mathrm{eff}/\Gamma_a^\mathrm{eff}$,
$n_1=P_1^\mathrm{eff}/\Gamma_1^\mathrm{eff}$, in terms of the
effective pumping rate and line broadenings
$P_a^\mathrm{eff}=P_a+\frac{Q_{a1}}{\Gamma_a+\Gamma_1}(P_a+P_1)$,
$\Gamma_a^\mathrm{eff}=\Gamma_a+Q_{a1}$,
$P_1^\mathrm{eff}=P_1+\frac{Q_{1}}{\Gamma_a+\Gamma_1}(P_a+P_1)$,
$\Gamma_1^\mathrm{eff}=\Gamma_1+Q_1$ and the Purcell exchange rate of
the cavity into the dot, $ Q_{a1}=4(g_1^\mathrm{eff})^2/\Gamma_1$, and
the dot into the cavity mode,
$Q_1=4(g_1^\mathrm{eff})^2/\Gamma_a$. The effective coupling strength
appearing in these expressions is given by
$g_1^\mathrm{eff}=g_1/\sqrt{1+\Big(\frac{2\Delta_1}{\Gamma_a+\Gamma_1}
  \Big)^2}$. By solving the system with two, three, etc., weakly
coupled dots, we find the general cavity effective parameters for $N$
dots:
\begin{subequations}
  \label{eq:MonNov2194824GMT2009}
  \begin{align}
    P_a^\mathrm{eff}=P_a&+\sum_{i=1}^N\frac{Q_{ai}}{\Gamma_i+\Gamma_a}(P_i+P_a)+\sum_{\{i,j\}}\frac{Q_{ai}Q_{aj}}{(\Gamma_a+\Gamma_i)(\Gamma_a+\Gamma_j)}(P_i+P_j+P_a)+\dots\,,\\
    \Gamma_a^\mathrm{eff}=\Gamma_a&+\sum_{i=1}^N Q_{ai}+\sum_{\{i,j\}}
    \frac{Q_{ai}Q_{aj}}{(\Gamma_a+\Gamma_i)(\Gamma_a+\Gamma_j)}(\Gamma_a+\Gamma_i+\Gamma_j)+\dots\,.
\end{align}
\end{subequations}
The sums are taken for increasingly large combinations of dots,
$\{i,j,k,\dots\}$, with $i<j<k<\dots$. They correspond to the exchange
of a photon between different dots. The larger the group, the smaller
its contribution to the effective parameters, as the weight is given
by the product $\alpha_i\alpha_j\alpha_k\dots$ where $\alpha_i\equiv
Q_{ai}/(\Gamma_a+\Gamma_i)$ is a dimensionless quantity, that is small
in the present model.

For simplicity, we consider that all QDs ($i=1,\dots, N$) are coupled
to the cavity mode with similar coupling strength, $g_i=g$, detunings,
$\Delta_i=\Delta$, that they have similar decay rates into the leaky
modes, $\gamma_i=\gamma$ and are excited at the same pumping rates
$P_i=P$ (then, $\Gamma_i=\Gamma$, $g_i^\mathrm{eff}=g^\mathrm{eff}$,
$Q_{ai}=Q_a$ and $\alpha_i=\alpha$). More realistically, there would
be a Gaussian distribution of these parameters, however, QDs with
higher effective coupling lead the dynamics, and this approximation
actually results in little loss of generality.  With this
simplification, we obtain the compact expressions:
\begin{subequations}
  \label{eq:MonNov2201251GMT2009}
  \begin{align}
    &P_a^\mathrm{eff}=P_a+\sum_{n=1}^N \frac{N!\alpha^n}{(N-n)! n!}(P_a+nP)=(P_a +N\frac{\alpha }{1+\alpha}P)(1+\alpha)^N\,,\\
    &\Gamma_a^\mathrm{eff}=\Gamma_a+ \sum_{n=1}^N \frac{N!\alpha^n}{(N-n)!
      n!}(\Gamma_a+n \Gamma )=(\Gamma_a +N\frac{\alpha }{1+\alpha}\Gamma)(1+\alpha)^N\,,
\end{align}
\end{subequations}

\begin{figure}[th]
  \centering
 \includegraphics[width=\linewidth,angle=0]{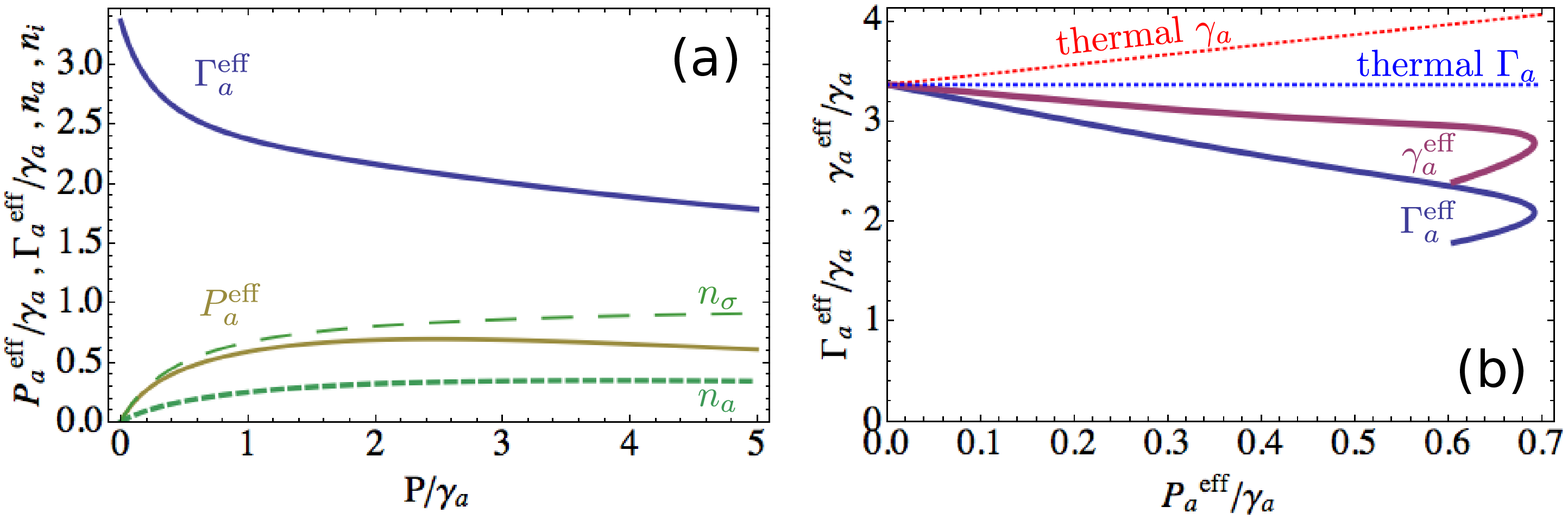}
 \caption{(Colour online) (a) Effective pumping rate,
   $P_a^\mathrm{eff}$ (solid brown) and cavity linewidth,
   $\Gamma_a^\mathrm{eff}$ (solid blue) as a function of the pumping
   rate of the dots, $P$. In dashed green, the dot populations,
   $n_\sigma$, quickly saturates to~$1$. In dotted green, the
   resulting cavity population, $n_a$. We checked that both
   populations agree with the Jaynes-Cummings numerical solution at
   higher orders. (b) $\Gamma_a^\mathrm{eff}$ (solid blue) decreases
   and $\gamma_a^\mathrm{eff}$ (solid purple) slightly decreases as a
   function of $P_a^\mathrm{eff}$ in contrast with a thermal
   excitation of the cavity where $\Gamma_a$ (dotted blue) remains
   constant and $\gamma_a$ (dotted red) increases linearly as a
   function of the same $P_a$. Note in (a) that $P_a^\mathrm{eff}$
   decreases from $P\approx 2.5$, which provokes the loop in (b) at
   $P_a^\mathrm{eff}\approx 0.7$. Parameters in this example are
   $N=15$, $g=0.3$, $\Delta=2$, $\gamma=0.5$, $P_a=0$. All quantities
   are in units of $\gamma_a$.}
  \label{fig:MonFeb16000429CET2009}
\end{figure}

For illustration, let us assume that the system is at zero temperature
and that the cavity is not excited directly, i.e., $P_a=0$. Then,
$P_a^\mathrm{eff}$ is fully produced by the Purcell emission of all
the weakly coupled QDs. In Fig.~\ref{fig:MonFeb16000429CET2009}, we
plot the effective cavity parameters as a function of the external QD
pumping $P$ (for some parameters given in the caption, typical of the
experiments).  All magnitudes are given in terms of the empty cavity
decay rate, $\gamma_a$. We can see in (a) that for $P<2.5$,
$P_a^\mathrm{eff}$ increases while $\Gamma_a^\mathrm{eff}$
monotonically decreases. If we plot one as function of the other, (b),
we find line narrowing with an increasing effective cavity
pumping. This is to be compared with a thermal excitation (dotted
lines), where the linewidth indeed does not change. The effective
decay rate $\gamma_a^\mathrm{eff}$ also decreases slightly, while it
would increase linearly with the cavity pumping in the thermal case.

It is not essential to invert the QD population (see $n_\sigma$ in
(a)) in order to obtain line narrowing. Similar results, with QD
saturation into $1/2$ instead of $1$, are obtained with a thermal
excitation of the QDs ($\gamma=\kappa+P$ and $\kappa=0.5$). The cavity
population remains quite low in any case (see $n_a$ in (a)), as
expected from weak-coupling, but this small contribution is enough to
cause qualitative differences in the strong coupling physics that
involve the QD of interest, $i=0$~\cite{delvalle09a}.

This simple model has room for arbitrary sophistication that can
relate $\Gamma_a^\mathrm{eff}$ and $P_a^\mathrm{eff}$ in
Eq.~(\ref{eq:WedApr7184035BST2010}) in all possible conceivable
ways. It is thus a shortcoming, in a configuration where the simplest
model displays opposite tendencies, to assume thermal
relationships between $\Gamma$s and $P$s~\cite{ridolfo10a,yao10a}, or,
for that matter, any particular constrain, such as
Eqs.~(\ref{eq:MonNov2201251GMT2009}). A more general view should be
adopted to let these parameters completely free and, based on
statistical inference, to extract correlations from them \emph{a
  posteriori}.

In conclusion, we have derived from a microscopic model of $N$ weakly
coupled and incoherently excited quantum dots, the effective cavity
pumping and decay rates for a master equation in the Lindblad
form. These are not linked by relationships of thermal equilibrium.
The QDs emit cavity photons via Purcell enhancement, providing a gain
medium for the cavity. As a result, the cavity spectral lineshape
narrows with increasing excitation, in contrast with a thermal
photonic excitation, where it remains constant.

\emph{Acknowledgements:} EdV acknowledges the support of a Newton
Fellowship.

\bibliographystyle{model1a-num-names}
\bibliography{Sci,books,cavfeed}

\end{document}